\documentclass[a4paper]{jpconf}
\usepackage{graphicx}
\usepackage{amsmath}
\usepackage[numbers]{natbib}   
\usepackage{hyperref}
\hypersetup{linkcolor=DarkGray}

\begin{document}
\begin{center}
 {\Large\bf  Progress in the Development of Multi-Element Monolithic Germanium Detectors in LEAPS-INNOV Project: Insights from Detector Performance Simulation \\[6mm]}

 {\small \it 
 N. GOYAL$^{1}$, S. APLIN$^{10}$, A. BALERNA$^{3}$, P. BELL$^{5}$, J. CASAS$^{11}$, M. CASCELLA$^{5}$, S. CHATTERJI$^{4}$, C. COHEN$^{7}$, E. COLLET$^{7}$, G. DENNIS$^{4}$, P. FAJARDO$^{7}$, E. N. GIMENEZ$^{4}$, H. GRAAFSMA$^{2}$, H. HIRESMANN$^{2}$, F. J. IGUAZ$^{1}$, K. KLEMENTIEV$^{5}$, T. KOLODZIEJ$^{8}$, L. MANZANILLAS$^{1}$, T. MARTIN$^{7}$, R. H. MENK$^{3,9}$, M. PORRO$^{10,12}$, M. QUISPE$^{11}$, B. SCHMITT$^{6}$, S. SCULLY$^{4}$, M. TURCATO$^{10}$, C. WARD$^{5}$, E. WELTER$^{2}$\\

 {\it $^1$ Synchrotron SOLEIL, L'Orme des Merisiers, Departementale 128, St-Aubin, France}		\\
  {\it $^2$ German Electron Synchrotron DESY, Hamburg, Germany}		\\
   {\it $^3$ The National Institute for Nuclear Physics INFN, Frascati, Italy}		\\
    {\it $^4$ Diamond Light Source Ltd, Oxfordshire, United Kingdom }		\\
     {\it $^5$ MAX IV Laboratory, Lund, Sweden}		\\
      {\it $^6$ The Paul Scherrer Institute PSI, Villigen PSI, Switzerland}		\\
       {\it $^7$ European Synchrotron Radiation Facility ESRF, Grenoble, France}		\\
  {\it $^8$ National Synchrotron Radiation Centre SOLARIS, Kraków, Poland}		\\
        {\it $^{9}$ Elettra Sincrotrone Trieste, Trieste, Italy}		\\
         {\it $^{10}$  European XFEL, Schenefel, Germany}		\\
         {\it $^{11}$  ALBA – CELLS Synchrotron, Cerdanyola del Vallès, Spain}		\\
 {\it $^{12}$  Department of Molecular Sciences and Nanosystems, Ca’ Foscari University of Venice, Italy}\\
 \textbf{email}: nishu.goyal@synchrotron-soleil.fr
 }
\end{center}

\begin{abstract}
This study presents a detailed simulation-based analysis of the detection limits of multi-element monolithic Germanium (Ge) detectors to cadmium traces in environmental soil samples. Using the capabilities of the Geant4 Monte Carlo toolkit in combination with the Solid State Detector Package, we evaluated the detection limit variation with the sample-to-detector distances and photon flux. These simulations were conducted to mimic realistic conditions, with a photon flux measured by the SAMBA beamline at the SOLEIL synchrotron facility. Our findings for the detection limit for trace amounts of pollutants in low concentrations like cadmium in the soil provide valuable insights for optimizing experimental setups in environmental monitoring and synchrotron-based applications, where precise detection of trace elements is critical.
\end{abstract}

\section{Introduction}
The development of advanced detection technologies is critical for enhancing the capabilities of synchrotron light sources and their applications. Among these technologies, multi-element monolithic Germanium (Ge) detectors have emerged as a promising solution for X-ray Absorption Fine Structure~(XAFS)\cite{9125956} and X-ray Fluorescence~(XRF) spectroscopy. These detectors are particularly suited for applications that require high sensitivity, excellent energy resolution, and the ability to discriminate between scattered and fluorescence photons, which is essential for accurate elemental analysis.
 This study focuses on the simulation of monolithic Ge detectors, developed under Work Package 2 (XAFS-DET)~\cite{Orsini2023} of the LEAPS-INNOV project, which aims to  push the performance of Ge detectors beyond the state-of-the-art to be able to compete with commercially available detectors for synchrotron applications. These detectors play a crucial role in achieving that goal by providing enhanced throughput and compactness while maintaining high-resolution performance.
 A significant challenge in X-ray spectroscopy, especially in environmental and geological studies, is the accurate detection of trace elements in complex matrices, such as soil samples. The detection limit is the parameter that defines the lowest concentration of an element that can be reliably measured~\cite{Iguaz_2022}. For this purpose, this study utilizes very versatile simulation tools, based on \textit{Geant4}~\cite{AGOSTINELLI2003250} and \textit{Solid state detectors}~\cite{Abt2021}, to model the response of these Ge detectors and estimate their detection limits under various experimental conditions.

The primary objective of this study is to take advantage of the validated full simulation chain to estimate the detection limit while varying the source-to-detector distance and photon flux.
\begin{figure}[h!]
    \begin{center}
        \includegraphics[scale=0.42,angle = 0,keepaspectratio]{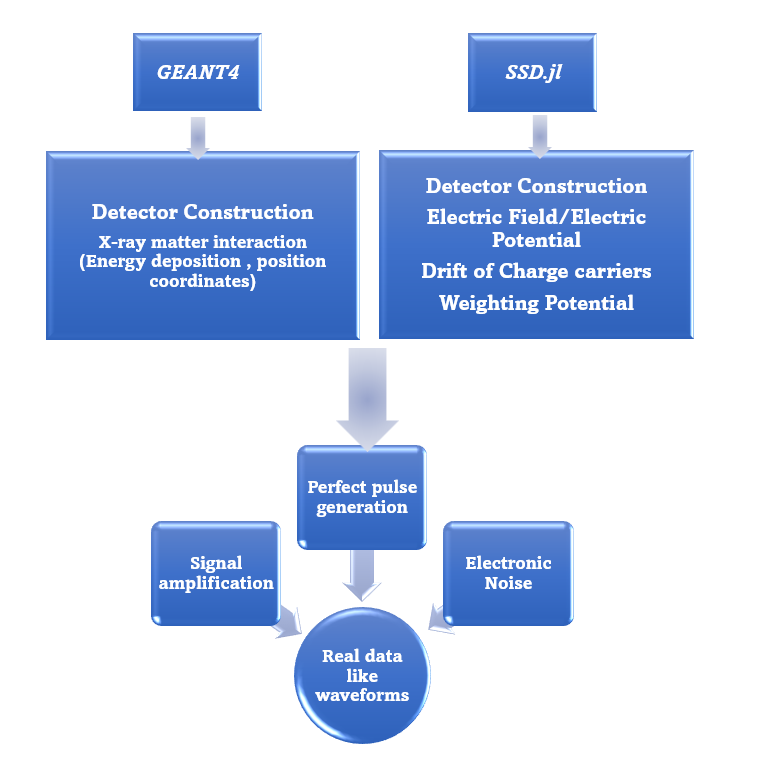}
              \end{center}
    \caption{ Flowchart illustrating the full simulation chain using Geant4 and SSD.jl, discussed in detail in the text}
\label{flowchart_simulation}
\end{figure}
\section{Detector Description and Features}
In the context of WP2, two multi-element monolithic germanium sensors, were fabricated by Mirion company, featuring two element sizes: 5~mm² dedicated for XAFS experiments and 20~mm² for XRF experiments, chosen to optimize throughput and detection efficiency respectively. The 5~mm² size significantly reduces from the typical 25~mm$^{2}$ used in synchrotrons, enabling future scalability with more channels. Both configurations share the same design, with the three outer elements dedicated to identifying the charge-sharing events for further electronic rejection, while the inner elements ensure uniform counting rates throughout the detector. The detector's multi-element configuration significantly enhances its throughput, making it suitable for high-count-rate applications typically required in synchrotron applications. The segmentation of sensor material and integration of Digital Pulse processor Xpress4~\cite{dennis2019first} also reduce charge-sharing effects, which is crucial for maintaining high energy resolution for varying input count rates. In addition to this, a newly developed full electronics chain by XGlab~\cite{xglab2023}, possessing front-end ceramics equipped with multichannel CMOS Cube preamplifier~\cite{6154396} and back-end ceramics for proper housing and connections for fast detector readout makes it more promising to reach a throughput/unit area from 50-250~kcps/mm². The compact design of the detector ensures its operation close to the sample, minimizing the path length of scattered photons, which is particularly important in applications like XAFS and XRF, where the ability to reject scattered radiation is key to improving the signal-to-background ratio and, consequently, the detection limit. Combining these characteristics makes the detector highly efficient for detecting a wide range of X-ray energies, extending up to 100~keV as required by many beamlines in facilities like  \textit{SOLEIL}, \textit{ESRF-EBS}, and \textit{DLS}.
\section{Description of the simulation chain}
The simulation of the detector's response was carried out using a combination of two very versatile Monte Carlo codes, \textit{Geant4}~\cite{AGOSTINELLI2003250} and \textit{SSD}~\cite{Abt2021} package to accurately model the interaction of X-rays with the Germanium detector and the subsequent charge transport, detector field calculation and energy reconstruction process~\cite{Manzanillas2023}.
Geant4 is widely used for simulating particle interactions with matter and was used to study the initial stages of X-ray interaction with the detector. A detailed model of the big pixel Ge detector, along with a Tungsten(W) collimator of thickness ~3mm and the sample of interest was defined within Geant4, allowing for precise tracking of the X-rays as they interact with the detector elements. The Geant4 output consists of the coordinates where the X-rays interact with the detector material and the corresponding energy deposited at each interaction point. These outputs are then fed into the SSD simulation. \\
The Julia programming-based SSD package is specifically designed for semiconductor detectors~\cite{Bezanson2012}. It is used to compute the electric field, diffusion of the charge carriers (e$^{-}$ $\&$ holes), and weighting potentials within the detector volume under an applied electric potential, see Fig.~\ref{flowchart_simulation}. \\
The electric field inside the detector is determined by various factors such as the geometry of the detector electrodes, bias voltage, and impurity density profile. In our case, a bias voltage of $\approx$ 200~V and a near-constant impurity density of around 10$^{10}$~cm$^{-3}$ were used. The drift of charge carriers depends on the electric field, temperature, and the drift direction relative to the Ge crystal axis. Charge carriers are collected by an electrode based on its position and the weighting potential associated with that electrode. The weighting potential is a theoretical concept that quantifies the charge fraction at a given position \textit{r} as seen by a specific contact \textit{C$_{i}$}. SSD applies the Shockley-Ramo theorem to calculate the time evolution of these induced charges on each electrode, which leads to the creation of noiseless, perfect pulses for each readout contact. Finally, realistic data like waveforms are generated by incorporating signal amplification using a simulated transfer function and electronic noise  (baseline data from the real experiment) into the simulated pulses. Using this approach with a background model allows us to predict the expected signal at a given photon flux and the sensitivity with respect to other influencing parameters like sample-to-detector distance and the pixel size. 
\begin{figure}[h!]
    \begin{center}
\includegraphics[scale=0.50,angle = 0,keepaspectratio]{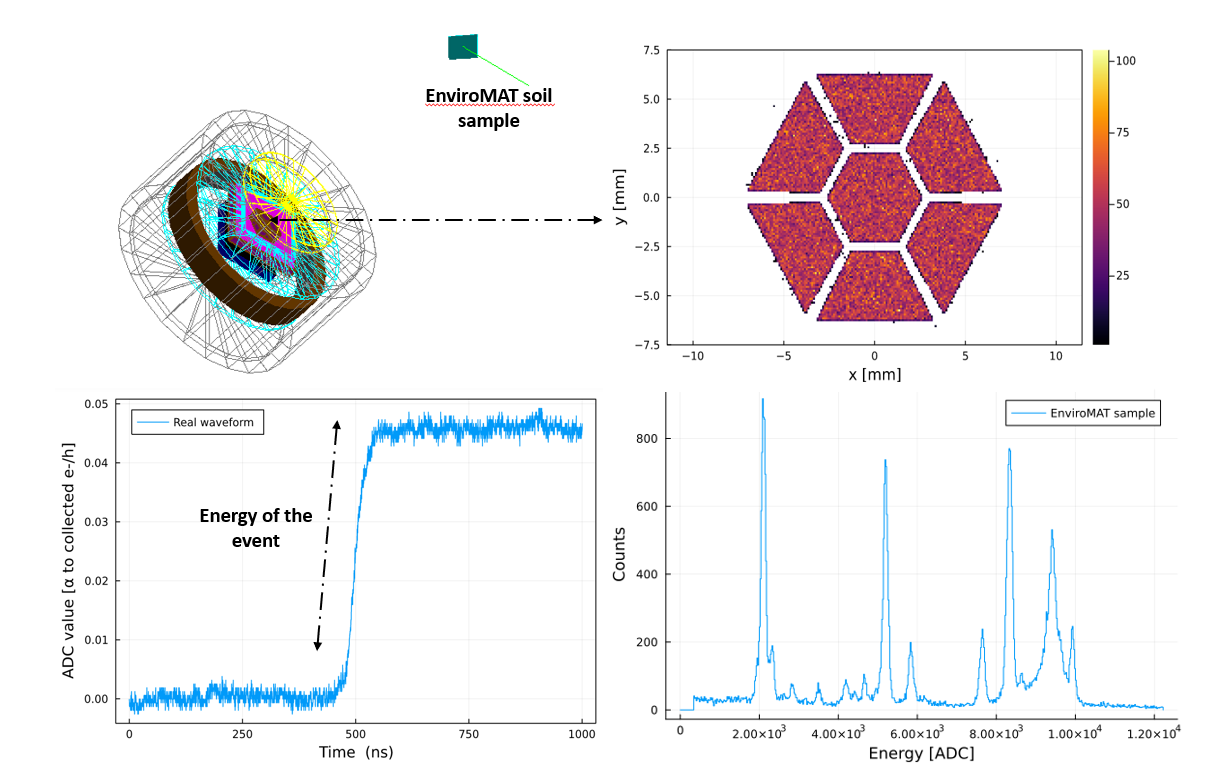}
    \end{center}
    \caption{Full simulation chain:     \textit{Top-left}: Geant4 constructed setup, which showcases the Ge detector with the big pixel along with a W collimator, and a 30~keV X-ray beam hitting the target sample; \textit{Top-right}: X-ray mapping for Ge  sensor produced using \textit{SSD} with Julia; \textit{Bottom-left}: simulated waveform for an event collected in a single contact; \textit{Bottom-right}:
     Simulated energy spectrum of \textit{EnviroMAT} soil sample using \textit{Geant4} output}
\label{geometryeventandenrgyspectra}
\end{figure}
\section{Estimation of Detection limit}
In this study, we have investigated the detection limit of the big pixel configuration of Ge detector to detect cadmium traces in a standard soil sample known as ENviroMAT\cite{enviromat}. 
 We explored how the detection limit for Cd varies with the sample-to-detector distance. The photon beam considered in these simulations (energy of $30~keV$, flux $\approx$ $3.47 \times 10^{10}$~ph/sec) was measured by a commercial 36-element HPGe detector on the \textit{SAMBA} beamline of the SOLEIL synchrotron facility to mimic experimental conditions.
 The simulation was performed by changing the samples to the detector distances in the Geant4 model, and we analyzed the reconstructed data using Julia codes to calculate the input parameters necessary, as mentioned in Eq.~\ref{dl_1} to quantify the minimum detectable concentration of \textit{Cd}. 

The detection limit to Cd at 90$\%$ C.L. is expressed as:

\begin{equation}
    DL(\text{ppm}) = \frac{3 \times C(\text{ppm})}{\sqrt{N_{\text{Pixels}} \times ICR_{\text{Pixel}} \times \left(1 - \frac{DT(\%)}{100.0}\right) \times \frac{S}{B} \times \frac{S}{T} \times T_{\text{exp}}}}
    \label{dl_1}
\end{equation}
\begin{equation}
{DL(\text{ppm}) = \frac{3 \times C(\text{ppm}) \times N_{\text{Bckgd.}}}{\sqrt{OCR_{\text{sig.}} \times {T_\text{exp}}}}}
\label{dl_2}
\end{equation}
           where,

           \begin{equation}
    OCR_{sig} = N_{pixels} \times \frac{ICR}{pixel} \times \left(1 - \frac{DT(\%)}{100}\right) \times \frac{S}{T}
    \label{detection_limit_2}
\end{equation}

\begin{figure}[h!]
    \begin{center}
\includegraphics[scale=0.45,angle = 0,keepaspectratio]{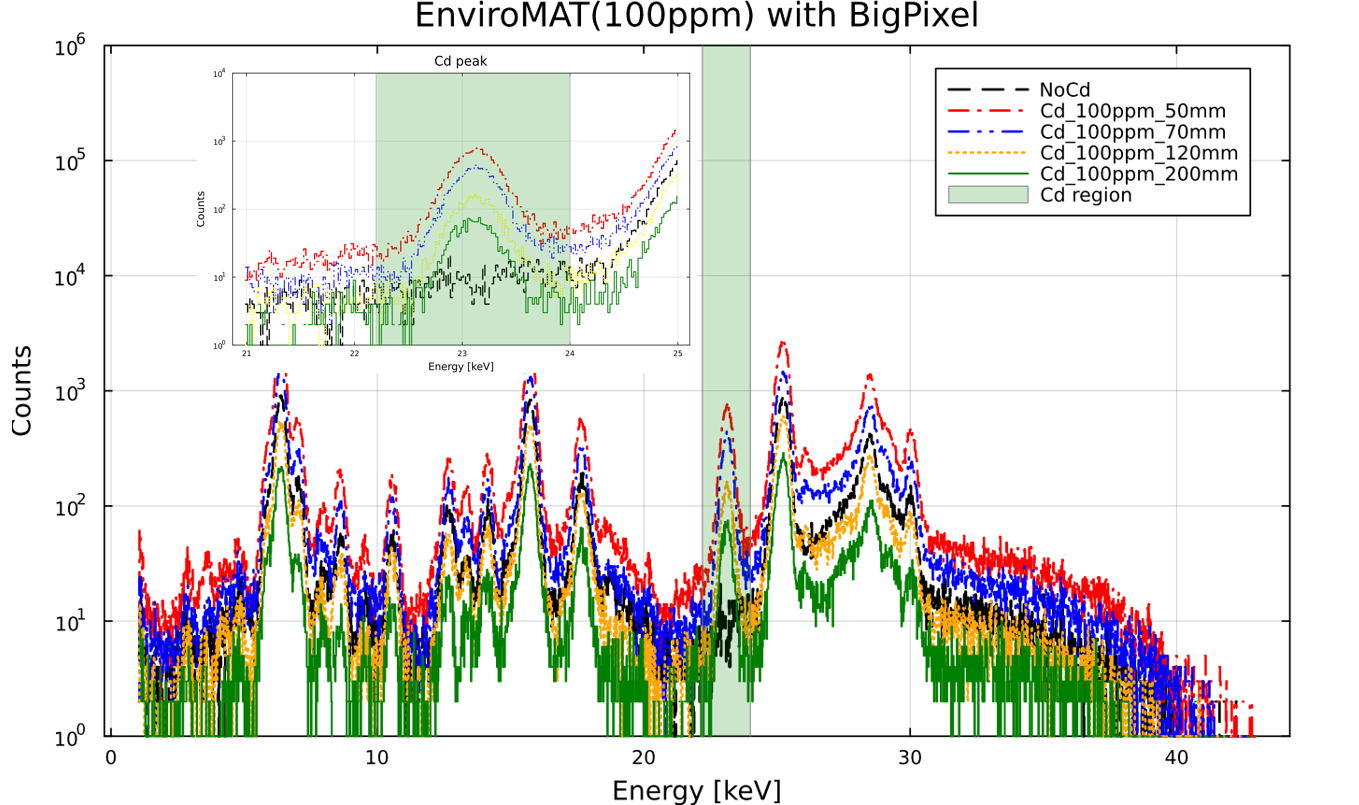}
    \end{center}
    \caption{Simulated element energy spectrum with Ge sensor placed at a distance of 50, 70, 120 and 200~mm with a soil sample of \textit{EnviroMAT} rich in \textit{Cd(}100~ppm) and without its trace. The exposure time considered here is 1 sec, with an incident beam energy of 30~keV, and the associated beam flux is $3.47 \times 10^{10}$~ph/sec. The shaded green area defines the
region of interest of the Cd K$\alpha$-peak.}
\label{EnviromatEnergyspectra}
\end{figure}

    \begin{itemize}
        \item $C$ is the concentration of the element (\textit{Cd}) in parts per million (ppm), 
        \item $N_{Bckgd}$ is the background count rate,
        \item $N_{pixels}$ is the total number of sensing elements
        \item $OCR_{signal}$ is the signal output count rate
        \item $ICR$ is the total input count rate
        \item $T_{exp}$ is the exposure time.
        \item $DT$ is the dead time of the digital pulse processor
        \item $S/B$ is signal-to-noise or signal-to-background ratio
        \item $S/T$ is signal to total spectrum ratio     
    \end{itemize}
The signal, representing the \textit{Cd} concentration of 100 ppm in the soil, and the background, representing a sample without it, are calculated within the region of interest (RoI) of the \textit{Cd} \textit{K$\alpha$}-peak in the energy range of 22.5-24~keV, as shown in Fig.~\ref{EnviromatEnergyspectra}. The total spectrum intensity is derived from the integral of the full energy spectrum, with the background intensity subtracted from the signal's integral to isolate the cadmium trace.

 The detection limit was evaluated for a photon beam of 30~keV over a range of distances, from 10 mm to 200~mm, and the beam flux intensity from 10$^7$ to 10$^{12}$~photons per second to determine how the geometry affects the detector’s sensitivity. The goal was to find the sweet spot where the signal-to-background ratio (SNR) is optimal, enabling the detector to identify even trace amounts of \textit{Cd} in soil. 
 
\par
\begin{figure}[h!]
    \begin{center}
\includegraphics[scale=0.30,angle = 0,keepaspectratio]{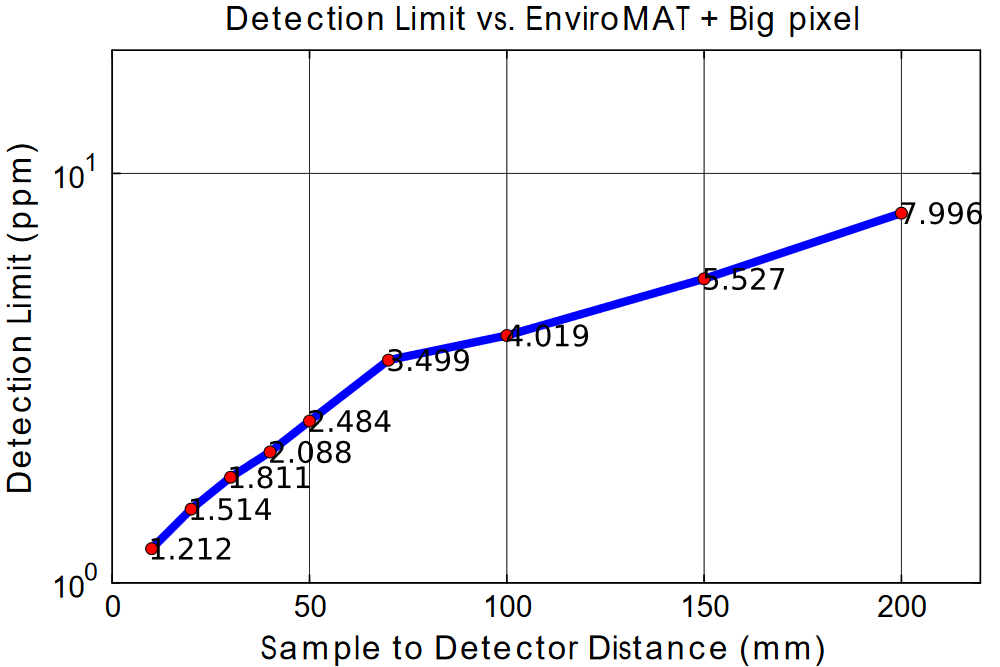}
    \end{center}
    \caption{ Detection limit (in ppm) as a function of sample to detector distance varying systematically from 10 to 200~mm exposed with a photon flux of $3.47 \times 10^{10}$~ph/sec.}
\label{Detctionlimitwithrespecttisampletodetectordistance}
\end{figure}
\begin{figure}[h!]
    \begin{center}
\includegraphics[scale=0.40,angle = 0,keepaspectratio]{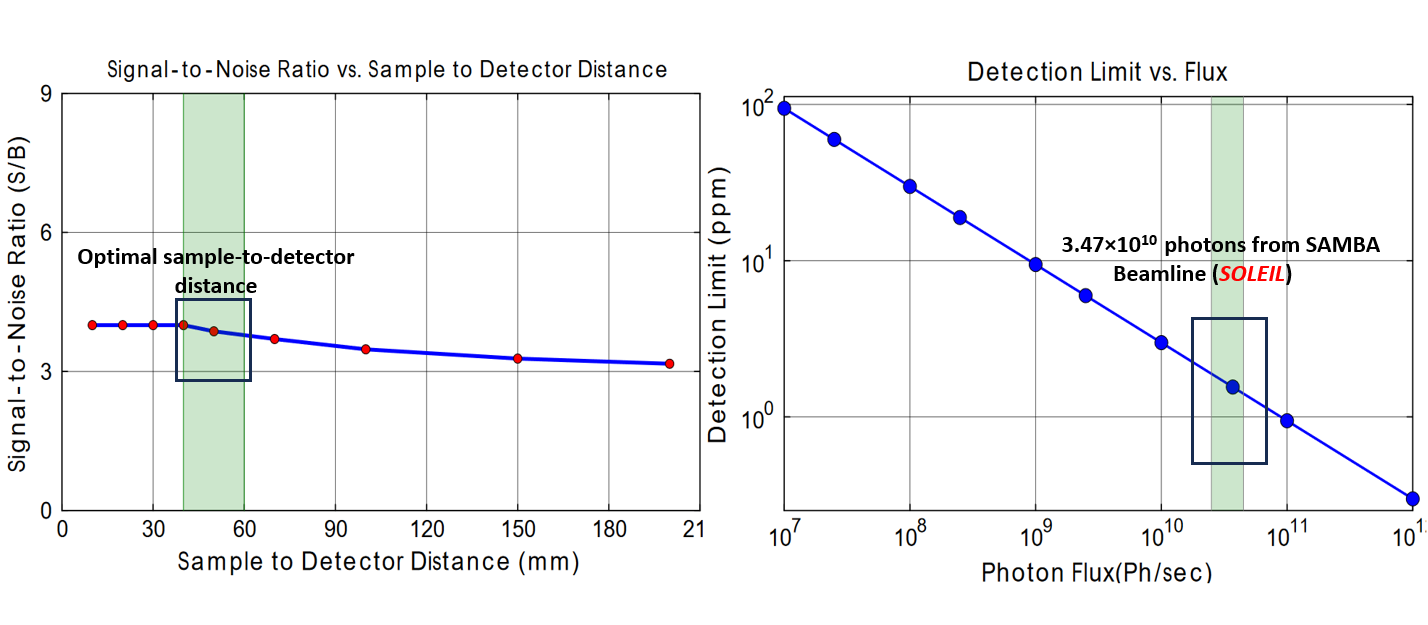}
    \end{center}
    \caption{(Left) Response of signal-to-noise ratio with respect to varying \textit{EnviroMAT} sample to detector distance in simulation Optimization of detection limits, analyzed over a S-D range of distances from 10 to 200~mm. (Right) The detection limit (in ppm) is represented on a logarithmic scale with respect to photon flux (ph/sec), improving with increasing number of incident photons}
\label{detectionlimitvsphotonflux}
\end{figure}
\section{Results and discussions}
The Fig.~\ref{Detctionlimitwithrespecttisampletodetectordistance} shows the detection limit (in ppm) as a function of the sample-to-detector distance (in mm) for the \textit{EnviroMAT} sample using the big pixel geometry. The detection limit, represented on a logarithmic scale, increases as the distance between the sample and detector increases. At shorter distances, for example, at 10~mm, the detection limit is low, around 1.21~ppm, indicating high sensitivity and the ability to detect smaller sample concentrations.
    As the distance increases, the detection limit rises gradually. At 50~mm, it reaches approximately 2.48~ppm, and by 100~mm, it reaches around 4.02~ppm.  This trend continues, with the detection limit reaching a value close to 7.99~ppm at 200~mm. This increase in the detection limit with distance implies that, as the detector moves farther from the sample, its ability to detect smaller concentrations decreases. The system's sensitivity reduces significantly as the distance between the sample and the detector increases.\\
    
This increase in the detection limit with distance can be attributed to two primary factors: attenuation of X-rays in air and increased susceptibility to noise at larger distances. As the sample-to-detector distance grows, a significant fraction of X-ray photons is absorbed or scattered in air before reaching the detector, resulting in a reduced number of photon interaction with the detector. This attenuation directly impacts the signal strength and, consequently, the detection limit. 
Additionally, the detector is  more susceptible to background  and electronic noise at larger distances, both of which degrade the signal-to-background ratio (S/B). These noise contributions were accounted for in our simulations using baseline  levels obtained from the experimental data. 

\section{Conclusion}
The LEAPS-INNOV detector consortium under their ambitious R$\&$D program commenced in April 2021 is focused on the development of state-of-the-art monolithic multi-element Ge detectors for synchrotrons.  Applications like XAFS, having the requirements of high throughput, and XRF, with energy-resolved spectroscopy grade detectors, can benefit. A versatile combination of Geant4 and SSD.jl was used to simulate the complete simulation chain to optimize the performance of the detectors under development.
 In this study, we have demonstrated the ability of Ge detectors to achieve low detection limits for cadmium in environmental soil samples. We have explored the impact of various factors, including detector-to-sample distance and beam flux intensity, through a comprehensive simulation approach on the detector's performance. The results show that our detector design is highly effective in distinguishing the signal from the background, even at lower concentrations.
\par

We have investigated how the sample-to-detector distance impacts the detection limit, and our simulations have shown that shorter distances can enhance the detection limit for smaller concentrations due to higher photon flux.
 We tested distances ranging from 10 to 200~mm and found that optimizing this distance is crucial for improving the detector’s sensitivity, particularly for the detection of trace elements. These results are significant for applications where precision is essential.
 In our case, we used the optimal 50~mm  sample-to-detector distance to evaluate the detection limit behavior under varying photon flux levels. The photon flux from the SAMBA beamline, which is marked in the simulation in Fig. \ref{detectionlimitvsphotonflux}, was used as a reference to analyze how the detection limit changes with different flux inputs. 
\par

Ultimately, the goal is to enhance the detection capabilities of monolithic Ge detectors, making them a valuable asset for applications like environmental monitoring, particularly in identifying trace elements in complex samples with high precision and sensitivity.
 Additionally, we aim to validate our fully developed simulation chain with the very first experimental data acquired with the first prototype of the monolithic Ge detector of the LEAPS-INNOV project, as anticipated to be scheduled shortly. This will aid us in fine-tuning the simulation, ensuring its suitability for a wide range of environmental and synchrotron-based applications.

\section*{Acknowledgement}
I would like to thank the Work Package 2 collaboration of LEAPS-INNOV project for their support and contributions to facilitate the progress so far. A special thanks to L. Manzanillas and F.J. Iguaz for their foundational work on the simulations and for guiding my efforts. I also appreciate the essential support of the computing services in using SUMO and TOPAZE IT clusters.
\par
\textit{This Project has received funding from the Horizon 202O research and programme of the European Union under grant number 101004728.}
\bibliographystyle{JHEP}
\bibliography{leaps}
\end{document}